\documentclass[12pt]{article}
\usepackage{amsmath,amssymb}
\usepackage{latexsym,graphicx,verbatim}

\usepackage[usenames]{color}

\def\6#1{{\underline{#1}}}
\def\m6#1{{\underline{#1}\,}}

\newdimen\Tdim
\def\ispan{{\setbox0=\hbox{i}%
\Tdim\ht0\advance\Tdim\dp0\rule[-\dp0]{0pt}{\Tdim}}}
\def\jspan{{\setbox0=\hbox{j}%
\Tdim\ht0\advance\Tdim\dp0\rule[-\dp0]{0pt}{\Tdim}}}
\def\Tspan#1{{\setbox0=\hbox{#1}%
\Tdim\ht0\advance\Tdim\dp0\advance\Tdim.55ex\rule[-\dp0]{0pt}{\Tdim}\box0}}

\def\be{\begin{eqnarray}}
\def\ben{\begin{eqnarray*}}
\def\ee{\end{eqnarray}}
\def\een{\end{eqnarray*}}
\def\Tr{{\rm Tr}}

\def\p{\partial}

\def\D{\mathcal{D}}

\def\=:{=\hspace{-.7em}\raisebox{1.1ex}{.}\hspace{.1em}\raisebox{-0.2ex}{.} }

\newcommand{\NF}{N_{\rm F}}
\newcommand{\NC}{N_{\rm C}}

\newcommand {\beq}{\begin{eqnarray}}
\newcommand {\eeq}{\end{eqnarray}}

\def\mn2changed#1{\textcolor{blue}{{\bf #1}}}

\makeatletter
\@addtoreset{equation}{section}

\renewcommand{\thefootnote}{\fnsymbol{footnote}}
\makeatother

\makeatletter
\newcommand{\thetablename}{Table}
\def\fnum@table{\thetablename\ \thetable}
\makeatother

\begin{document}
\thispagestyle{empty}
\begin{flushright}
RIKEN-TH-165\\
{\tt arXiv:} \\
August, 2009 \\
\end{flushright}
\vspace{3mm}
\begin{center}
{\LARGE  
Color Magnetic Flux Tubes in Dense QCD.\\
II: Effective World-Sheet Theory
} \\ 
\vspace{20mm}

{\normalsize
Minoru~Eto$^{a}$, 
Eiji~Nakano$^b$ and
Muneto~Nitta$^c$}
\footnotetext{
e-mail~addresses: \tt
meto(at)riken.jp;
e.nakano(at)gsi.de;
nitta(at)phys-h.keio.ac.jp.
}

\vskip 1.5em
{\footnotesize
$^a$ {\it Theoretical Physics Laboratory, RIKEN, 
Saitama 351-0198, Japan
}
\\
$^b$ {\it 
Extreme Matter Institute, GSI, Planckstr. 1, D-64291 Darmstadt, Germany
}
\\
$^c$ 
{\it Department of Physics, and 
Research and Education Center for Natural Sciences,
Keio University,\\ 
4-1-1 Hiyoshi, Yokohama,
Kanagawa 223-8521, Japan
}
}
 \vspace{12mm}

\abstract{

Color magnetic flux tubes appear 
in the color-flavor locked  phase of high density QCD, 
which exhibits color superconductivity as well as 
superfluidity. 
They are non-Abelian superfluid vortices and 
are accompanied by  
orientational zero modes in the internal space  
associated with 
the color-flavor locked symmetry 
spontaneously broken 
in the presence of the vortex. 
We show that those zero modes are localized 
around the vortex 
in spite of the logarithmic divergence of its tension,  
and derive the low-energy effective theory 
of them on the world-sheet of the vortex-string. 
}

\end{center}

\vfill
\newpage
\setcounter{page}{1}
\setcounter{footnote}{0}
\renewcommand{\thefootnote}{\arabic{footnote}}


\section{Introduction}

It seems likely from theoretical studies that color superconducting phase 
exists in high density and low temperature region of QCD phase diagram 
\cite{Rajagopal:2000wf}. 
The color superconducting phase is classified roughly 
into so-called 2SC \cite{Rapp:1997zu,Alford:1997zt}
and color-flavor locked (CFL) \cite{Alford:1998mk} phases depending on the number of flavor 
participating in a condensation, 
and further into many other variants in more realistic situations with 
color and charge neutrality, effects of the chiral anomaly, and finite quark masses 
\cite{Alford:2007xm,Rajagopal:2000ff}. 
Such a state of matter is considered to be realized in the core of compact stars, 
or during the evolution after collision experiments. 
To capture their signatures, 
it is necessary to figure out various properties of color superconductivity. 

In three flavor case which we are interested in, and in higher density regions where 
effects of the quark masses can be ignored and the three flavor symmetry effectively hold, 
the CFL phase would take place with the order parameter: 
\beq
\Phi_{k \gamma}^{L(R)}= \epsilon_{ijk}\epsilon_{\alpha\beta\gamma} 
\langle q_{i\alpha}^{L(R)} C 
q_{j\beta}^{L(R)}\rangle \propto \delta_{k \gamma}, 
\eeq 
where $i,j,k$ and $\alpha,\beta,\gamma$ are flavor and color 
indices. 
This diagonal configuration which locks flavor and color, minimizes the free energy 
\cite{Alford:1998mk}. 
In the CFL phase the symmetry 
$G\simeq SU(3)_{\rm C} \times SU(3)_{\rm L} \times SU(3)_{\rm R} 
\times U(1)_{\rm B}$ 
breaks down to the diagonal one $H\simeq SU(3)_{\rm C+L+R}\equiv SU(3)_{\rm C+F}$, 
where we consider massless case and left- and right-handed quarks are separated. 
The Higgs mechanism provides masses of all eight gluons,\footnote{
In realistic situation with electro-magnetic gauge field, 
seven gluons and one linear combination of the 8th gluon and electro-magnetic photon acquire Higgsed masses. 
The other orthogonal combination remains massless.
In the present work, however, we ignore the electro-magnetism for simplicity.} 
and there appear eight Nambu-Goldstone (NG) bosons 
(the CFL mesons) associated with chiral symmetry breaking. 
Also, spontaneous breaking of the baryon number symmetry $U(1)_{\rm B}$ 
generates a phonon as the associated NG boson. 
Low-energy effective theories of the CFL mesons 
and the $U(1)_{\rm B}$ phonon
have been derived in \cite{Casalbuoni:1999wu} 
and \cite{Son:2002zn}, respectively.

When a symmetry of a system is spontaneously broken in 
the ground state, 
there appear various kind of
topological defects, 
corresponding to a non-trivial topology of the order parameter space.
Particular attentions have been paid to vortices 
determining the dynamics of 
a system with spatially rotating and/or 
under an external magnetic field,  
such as pulsars with strong magnetic field. 
Therefore, in this paper, we study vortices in the CFL phase 
\cite{Giannakis:2001wz}--\cite{Eto:2009kg}.
In such systems one might expect that 
there appear stable vortices associated 
with $U(1)_{\rm B}$ symmetry breaking, 
but this is only true for the confining phase like a hadronic matter. 
The $U(1)_{\rm B}$ superfluid vortices
appear also in the deconfining CFL phase 
as a response to rotation \cite{Forbes:2001gj,Iida:2002ev},
but each of them is unstable to 
decay into three non-Abelian vortices 
found in \cite{Balachandran:2005ev} 
which are color magnetic flux tubes. 
This is because 
the total tension of the three well-separated non-Abelian vortices 
is 1/3 of that of one $U(1)_{\rm B}$ vortex. 
This decay is inevitable because of 
a long range repulsive force between non-Abelian vortices 
\cite{Nakano:2007dr}. 
Moreover it has been found in \cite{Eto:2009kg} that
one non-Abelian semi-superfluid vortex carries 
1/3 amount of the color flux of the color magnetic vortex studied in
\cite{Giannakis:2001wz,Iida:2002ev,Iida:2004if}. 
These vortices are called semi-superfluid vortices 
which respond not only to color field but also to rotation 
like superfluid vortices \cite{Balachandran:2005ev}. 
The non-Abelian vortex is therefore the most fundamental vortex 
in the CFL phase,  
which is topologically stable and has the minimum tension and flux. 
In the core of rotating stars which exhibits the CFL phase 
the non-Abelian vortices might form a vortex lattice.\footnote{
It has been suggested in \cite{Shahabasyan:2009zz} that 
the oscillations propagate
in the plane perpendicular to vortex-strings 
in a vortex lattice.} 
Its lattice structure can be determined 
by details of vortex-vortex interaction. 
It should be noted here that 
as the most significant feature of the non-Abelian vortices, 
there appear, around the vortex-string, further NG zero modes
associated with the additional symmetry breaking 
due to the advent of the vortex: 
$H=SU(3)_{\rm C+F} \rightarrow K
=[U(1)\times SU(2)]_{\rm C+F}$. 
These modes parametrize the complex projective space 
$H/K=\mathbb{C}P^{2}$ \cite{Nakano:2007dr}, 
and they are called orientational zero modes. 
Points in $\mathbb{C}P^{2}$  correspond one-to-one to color 
degrees of freedom which the vortex carries.

In the previous paper \cite{Eto:2009kg} 
we have constructed full numerical solutions of 
the semi-superfluid non-Abelian vortices 
with diverse choices of parameters. 
We have analytically shown that both the scalar and gauge fields 
asymptotically behave as $e^{- m r}$ 
with $m = {\rm min}(m_G,m_\chi)$, 
where $m_G$ and $m_{\chi}$ are the masses of 
the gluons and the traceless part of the scalar fields, respectively.
We also have numerically evaluated 
the width of the color flux and found that 
it is not always the penetration depth, the Compton wave 
length $m_G^{-1}$. 
When the gluon mass is smaller than the scalar masses 
the width cannot become larger than certain values 
determined by the masses of other fields, 
so we have found that the color flux 
is enforced to reside in the scalar core.

The orientational zero modes $\mathbb{C}P^{N-1}$ of non-Abelian vortices 
were 
first found in the context of supersymmetric $U(N)$ QCD
\cite{Hanany:2003hp} in which 
$U(1)_{\rm B}$ is also gauged, see \cite{Tong:2005un} as a review.  
The non-Abelian vortices appearing in these theories 
are local vortices which have finite tension 
and are at critical coupling 
(called Bogomol'nyi-Prasad-Sommerfield (BPS) states 
in the context of supersymmetry).
Thanks to supersymmetry,  
the normalizability of the orientational zero modes $\mathbb{C}P^{N-1}$ 
was proved and 
the 1+1 dimensional $\mathbb{C}P^{N-1}$ model 
with a suitable decay constant (overall constant) 
was obtained as the effective world-sheet theory 
of the non-Abelian vortex, for instance \cite{Gorsky:2004ad}.
On the other hand, 
in the case of our non-Abelian semi-superfluid vortex, 
the normalizability of 
the orientational zero modes 
has not been shown yet. 
We have only shown in \cite{Nakano:2007dr} 
that the orientational zero modes do not affect boundary 
condition, which is necessary but not sufficient 
for the normalizability.\footnote{
Non-Abelian global vortices appear 
in the chiral symmetry breaking, 
where all symmetries are global \cite{Balachandran:2002je}.
In this case, 
the corresponding ${\mathbb C}P^2$ zero modes are 
obviously non-normalizable because they change the boundary condition,
and therefore cannot be regarded as zero modes associated
with the vortex itself. 
Those vortices also appear in QCD at very high density in which 
$U(1)_{\rm A}$, originally broken by instantons, 
is approximately recovered.
}
For instance the orientational zero modes 
of non-Abelian semi-local vortices have been shown to be non-normalizable 
(if the size moduli are non-zero) \cite{Shifman:2006kd}, 
although those modes do not affect boundary conditions.
The question whether orientational zero modes 
of the non-Abelian semi-superfluid vortex 
are normalizable or not  
remains as a significant problem in order to study 
its dynamics.

In the present work 
we explicitly show the normalizability of the orientational zero modes and 
derive the low-energy effective world-sheet theory 
of a non-Abelian semi-superfluid vortex. 
To this end,
we generalize the derivation of the effective action of 
the BPS non-Abelian vortex-string by 
Gorsky, Shifman and Yung \cite{Gorsky:2004ad} 
where the decay constant (overall constant) of 
the $\mathbb{C}P^{N-1}$ model was found to be $4 \pi/g_s^2$ with 
a gauge coupling constant $g_s$. 
For our case of a non-Abelian vortex in the CFL phase 
we find that the decay constant of the $\mathbb{C}P^{2}$ model
does not coincide with $4 \pi/g_s^2$ of the BPS case. 
It can be larger or smaller depending on the parameter regions. 
Our work will be the first step to study dynamics 
of semi-superfluid non-Abelian vortex strings 
which will be relevant for instance in 
the neutron star physics.

In the case of Abelian vortex-strings, 
only translational zero modes are localized around
a vortex,
and the dynamics of a single vortex-string 
is described by the Nambu-Goto action
\beq
 S = - T \int d^2 \sigma \sqrt{ - \gamma}
\eeq
with the tension $T$ and the induced metric $\gamma$ on the world-sheet. 
It is well known that Kelvin waves 
propagates along a vortex-string. 
In our case, this is complemented by 
the $\mathbb{C}P^{2}$ model action. 
These two kinds of modes arize for instance at finite temperature.

\begin{figure}[ht]
\begin{center}
\includegraphics[width=7cm]{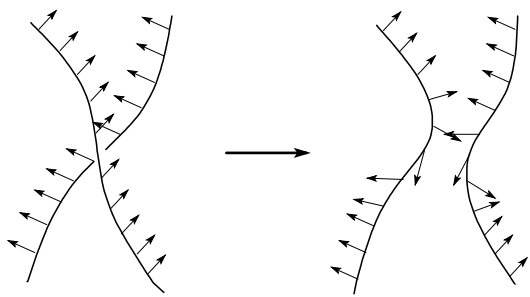}
\caption{Schematic picture of the reconnection of two 
non-Abelian vortex strings.
The allows stand for the $\mathbb{C}P^{N-1}$ orientations.
}
\end{center}
\label{fig:reconnection}
\end{figure}

On the other hand, 
dynamics of multiple vortices 
such as reconnection of two vortex-strings 
have been studied in various area from condensed matter 
physics to cosmology \cite{Vilenkin}. 
For instance when two vortex-strings reconnect with each other 
in a helium superfluid, 
the Kelvin waves are induced and this process
is considered to play an essential role 
in quantum turbulence. 
In the case of non-Abelian vortices, 
the reconnection of BPS local non-Abelian vortex-strings 
was studied in \cite{Eto:2006db}. 
It was found that 
even if two non-Abelian vortex-strings 
initially have different ${\mathbb C}P^{N-1}$ orientations 
in the internal space, 
their orientations must be aligned at the collision point 
and that the reconnection always occurs 
as in Fig.~\ref{fig:reconnection}.
We expect the same thing occurs in 
the collision of two semi-superfluid non-Abelian vortices.
When two non-Abelian semi-superfluid vortex-strings reconnect, 
it is expected that not only the Kelvin waves 
but also waves in the internal $\mathbb{C}P^{2}$ space arise 
(see right of Fig.~\ref{fig:reconnection}).
This may induce a new kind of turbulence or entangled network of 
non-Abelian strings, which is different 
from a helium superfluid.
Our work also provides a basis to proceed with 
more applicative studies on dynamics of multiple vortex systems, 
such as a vortex lattice 
in the core of a neutron star.

This paper is organized as follows.
In Sec.~\ref{sec:2} we provide 
the basic ingredients for our calculations,
the time-dependent Landau-Ginzburg Lagrangian, 
the non-Abelian semi-superfluid vortex, 
and its orientational zero modes in the color-flavor space. 
In Sec.~\ref{sec:3} we construct 
the low-energy effective theory 
for the orientational zero modes of 
a semi-superfluid vortex-string. 
Sec.~\ref{sec:4} is devoted to conclusion and discussion.

\section{Color Magnetic Flux Tubes} \label{sec:2}

We start with a Ginzburg-Landau effective Lagrangian 
for the CFL order parameters 
$\Phi^L$ and $\Phi^R$. 
Since at high density region a perturbative calculation shows
mixing terms between $\Phi^L$ and $\Phi^R$ are negligible, 
we simply assume $\Phi^L=\Phi^R \equiv \Phi$, 
and fix their relative phase to unity 
\cite{Balachandran:2005ev}. 
Then the static Ginzburg-Landau Lagrangian has been obtained 
as a low energy
effective theory of the high density QCD in the CFL phase 
\cite{Iida:2000ha,Giannakis:2001wz} 
\beq
{\cal L}^{(1)} = \Tr\left[- \frac{1}{4} F_{mn}F^{mn} 
+ K_1 \D_m\Phi^\dagger\D^m \Phi 
- \lambda_2 (\Phi^\dagger\Phi)^2 - n^2 \Phi^\dagger\Phi 
\right]
- \lambda_1\left(\Tr[\Phi^\dagger\Phi]\right)^2 
\label{eq:lsm}
\eeq
where $\D_m = \p_m - i g_s A_m$, 
$F_{mn} = \p_m A_n - \p_n A_m - i g_s\left[A_m,A_n\right]$ 
with the spatial indices $m,n=1,2,3$, 
and $\Tr [T^aT^b] = \delta^{ab}$ with color indices $a=1,2,\cdots,8$. 
Here $g_s$ is $SU(3)_{\rm C}$ gauge coupling constant.
In addition to the Lagrangian (\ref{eq:lsm}),  
the time-dependent Ginzburg-Landau Lagrangian 
contains\footnote{
We mimic the time-dependent Ginzburg-Landau Lagrangian 
known in the conventional superconductors \cite{Larkin}.
We neglect terms like $\partial_0 \Tr [(\Phi^\dagger \Phi)^n]$
and $\partial_0 [\Tr (\Phi^\dagger \Phi)]^n$ with $n \geq 2$ because 
we regard them as higher order terms containing 
the fourth order of fields and a time derivative. 
Even if one includes these terms in the Lagrangian, our results in Sec.~\ref{sec:3} 
are not changed.
}
\beq
{\cal L}^{(0)} &=& \Tr\left[- \frac{1}{2} F_{0m}F^{0m} 
+ K_0 (\tilde \D_0 \Phi)^\dagger \tilde \D^0 \Phi
\right],
\eeq
with
\beq
\tilde D_0 \Phi &\equiv& (\D_0 - 2 i \alpha)\Phi.
\eeq
The full Lagrangian ${\cal L} = {\cal L}^{(1)} + {\cal L}^{(0)}$ respects the $SO(3)$ spatial rotation,
the $SU(3)_{\rm C}$ gauge symmetry 
and the $SU(3)_{\rm F}$ flavor symmetry. 
While the parameters $K_1,m,\lambda_1,\lambda_2$ in 
the static Lagrangian (\ref{eq:lsm}) have been
obtained in a weak coupling regime of high density QCD 
\cite{Iida:2000ha,Giannakis:2001wz}, 
the parameters $K_0$ and $\alpha$ 
in a time dependent Lagrangian 
have not yet been determined microscopically in the literature 
to our knowledge.
However they must be determined in principle from 
the microscopic QCD Lagrangian. 
In general  $\alpha=\alpha_1+i \alpha_2$ is a complex function and 
is related to medium effects. 
Thus as one goes to QCD vacuum where 
both of temperature and the baryon-number density are zero, 
the function $\alpha$ vanishes to restore the Lorentz invariance. 
Instead of deriving the unknown parameters $K_0$ and $\alpha$
in ${\cal L}^{(0)}$ from QCD, we leave them as free parameters 
in this paper.
We can decompose the time covariant derivative as 
$\Tr[\tilde \D_0 \Phi^\dagger \tilde \D^0 \Phi] = 
\Tr[\D_0\Phi^\dagger\D^0\Phi + \alpha_1 j_0 + \alpha_2 \tilde{j}_0 + |\alpha|^2 \Phi^\dagger\Phi]$ with
$j_0 \equiv i[(\D_0\Phi)^\dagger \Phi - \Phi^\dagger \D_0\Phi]$ and 
$\tilde{j}_0 \equiv (\D_0\Phi)^\dagger \Phi + \Phi^\dagger \D_0\Phi = \p_0\left(\Phi^\dagger\Phi\right)$.  
Thus our Lagrangian 
is generic in a sense that it consists of all possible 
$SO(3) \times SU(3)_{\rm C}\times SU(3)_{\rm F}$ 
invariant terms in an expansion of  
the time and space derivatives 
and the order parameter $\Phi$ 
up to quadratic order in total. 
It is convenient to write the full Lagrangian 
\beq
{\cal L} &=& 
\Tr\left[- \frac{1}{4} F_{\mu\nu}F^{\mu\nu} 
+ K_0 (\D_0\Phi^\dagger\D^0 \Phi + \alpha_1 j_0 
+ \alpha_2 \tilde{j}_0) + K_1 \D_m\Phi^\dagger\D^m \Phi \right] - V,
\label{eq:full_lag}\\
V &=& \Tr \left[
\lambda_2 (\Phi^\dagger\Phi)^2 - m^2 \Phi^\dagger\Phi 
\right]
+ \lambda_1\left(\Tr[\Phi^\dagger\Phi]\right)^2,
\eeq
with $m^2 \equiv - n^2 + |\alpha|^2$. 
For the stability of the ground state, 
we consider the parameter region 
$m^2 > 0$, $\lambda_2 > 0$ and $3\lambda_1 + \lambda_2 > 0$.

The action of color, flavor and baryon symmetries on $\Phi$ 
is given by 
\beq 
 \Phi \to e^{i \theta} U_{\rm C} \Phi U_{\rm F},
 \quad U_{\rm C} \in SU(3)_{\rm C}, \; 
 U_{\rm F} \in SU(3)_{\rm F}, \;
 e^{i \theta} \in U(1)_{\rm B} .  \label{eq:action}
\eeq
There is some redundancy of the action of these symmetries.
The actual symmetry is given by 
\beq
 G  \equiv 
    \frac{SU(3)_{\rm C} \times SU(3)_{\rm F} \times U(1)_{\rm B}}
   {(\mathbb{Z}_3)_{\rm C+B} \times (\mathbb{Z}_3)_{\rm F+B}},
\eeq
where the discrete groups in the denominator
do not change $\Phi$ and are removed from 
$G$ \cite{Balachandran:2005ev,Eto:2009kg}.

By using the symmetry $G$, one can choose
a vacuum expectation value (VEV) as
\beq
\left<\Phi\right> = v {\bf 1}_3,\qquad
v^2 \equiv \frac{m^2}{2(3\lambda_1 + \lambda_2)} > 0 
\label{eq:vev}
\eeq
without loss of generality. 
By this condensation the gauge symmetry $SU(3)_{\rm C}$ 
is completely broken, 
and the full symmetry $G$ is spontaneously broken down to
\beq
H 
= \frac{SU(3)_{\rm C+F}}{(\mathbb{Z}_3)_{\rm C+F}} 
  \label{eq:H}.
\eeq
Therefore the order parameter space 
(the vacuum manifold) is given by
\beq 
 M \simeq G/H 
 = \frac{SU(3)_{\rm C-F} \times U(1)_{\rm B} }
 {(\mathbb{Z}_3)_{\rm C-F+B}} 
 = U(3).
\eeq
This space is parameterized by $SU(3)$ would-be NG bosons, 
which are eaten by eight gluons,  
and one massless NG boson of the spontaneously broken $U(1)_{\rm B}$.
The mass spectra around the Higgs ground state (\ref{eq:vev}) 
can be found by perturbing $\Phi$ as
\beq
\Phi = v {\bf 1}_3 
 + \frac{\phi + i \varphi}{\sqrt 2} {\bf 1}_3 
 + \frac{\chi^a + i \zeta^a}{\sqrt{2}}T^a.
\eeq
The trace part $\phi$ and $\varphi$ belong to the singlet of 
the color-flavor locked symmetry
whereas the traceless part $\chi$ and $\zeta^a$ belong to
the adjoint representation of it.
The gluons get mass 
with eating $\zeta^a$ by the Higgs mechanism. 
The masses of fields are given by 
\beq
m_G^2 = 2g_s^2v^2 K_1,\quad
m_{\phi}^2 = \frac{2m^2}{K_1},\quad
m_{\varphi}^2 = 0,\quad
m_{\chi}^2 = \frac{4\lambda_2 v^2}{K_1},
\eeq
where $m_G$ is the mass of the $SU(3)$ massive gluons 
and $\varphi$ is the NG boson (phonon) associated
with the spontaneously broken $U(1)_{\rm B}$ symmetry.
The trace part $\phi$ and 
the traceless part $\chi$ of $\Phi$ are massive bosons.

Let us construct a minimal vortex solution in the CFL phase.
We make the standard ansatz for a static vortex-string configuration 
parallel to the $x_3$-direction
(perpendicular to the $x_1$-$x_2$ plane):
\beq
\Phi(r,\theta) &=& v\, {\rm diag}\left(
e^{i\theta}f(r),\ g(r),\ g(r)\right),\\
A_i(r,\theta) &=& \frac{1}{g_s} \frac{\epsilon_{ij}x^{j}}{r^2}
 \left[1-h(r)\right] 
{\rm diag}
\left(
-2/3,\ 1/3,\ 1/3
\right),
\label{eq:minimum-winding}
\eeq
with $i,j=1,2$. 
Equations of motion for the profile function $f(r),g(r),h(r)$ are of the form
\beq
&&f''+\frac{f'}{r}
-\frac{(2 h+1)^2}{9 r^2}f-\frac{m_{\phi }^2}{6} f \left(f^2+2 g^2-3\right) 
-\frac{m_{\chi }^2}{3} f \left(f^2-g^2\right) = 0,
\label{eq:1}\\
&&g''+\frac{g'}{r}
-\frac{(h-1)^2}{9 r^2}g-\frac{m_{\phi }^2}{6} g \left(f^2+2 g^2-3\right) 
+\frac{m_{\chi }^2}{6} g \left(f^2-g^2\right) =0,
\label{eq:2}\\
&&h''-\frac{h'}{r} - \frac{m_G^2}{3}  \left(g^2 (h-1)+f^2 (2 h+1)\right) = 0.
\label{eq:3}
\eeq
We solve these differential equations with the following boundary conditions
\beq
\left\{
\begin{array}{cl}
(f,g,h) \to (1,1,0) \quad &\text{as}\quad r \to \infty,\\
(f,g',h) \to (0,0,1) \quad &\text{as}\quad r \to 0.
\end{array}
\right.
\label{eq:bc}
\eeq
For the regularity of the field $\Phi$, the profile function 
$f(r)$ must vanish at the origin. This means that in the center of the flux
tube, there exists an ungapped component. 
An approximate numerical solution with $g=1$ was first obtained in 
\cite{Balachandran:2005ev}.
The full numerical solution without any approximation 
has been recently obtained by the relaxation method 
in diverse choices of 
parameters \cite{Eto:2009kg}.

The above ansatz can be written in the different way 
\beq
\Phi(r,\theta) &=& v\,
e^{i\theta\left(\frac{1}{\sqrt{3}}T_0-\sqrt{\frac{2}{3}}T_8 \right)}
\left(
\frac{F(r)}{\sqrt 3} T_0 - \sqrt{2 \over 3} G(r) T_8
\right) \label{eq:ans_s}
\\
A_i(r,\theta) &=& \frac{1}{g_s} \frac{\epsilon_{ij}x^{j}}{r^2}
 \left[1-h(r)\right] \sqrt{\frac{2}{3}} T_8 
\label{eq:ans_a}
\eeq 
with new fields
\beq
F \equiv f + 2g,\quad G \equiv f-g,
\eeq
and the $U(3)$ generators
\beq
 T_0 = \frac{1}{\sqrt{3}}{\rm diag}(1,1,1),\quad
 T_8 = \frac{1}{\sqrt{6}}
 {\rm diag}(-2,1,1) .
\eeq
The piece proportional to $T_8$ in $\Phi(r,\theta)$ breaks the color-flavor locked symmetry
$H=SU(3)_{\rm C+F}$ down to $K = U(2)_{\rm C+F}$. 
This yields the NG modes (the orientational zero modes)
\beq
\frac{H}{K} = \frac{SU(3)}{SU(2)\times U(1)} \simeq \mathbb{C}P^2.
\eeq

\section{Low-Energy Effective World-Sheet Theory}\label{sec:3}

The semisuperfluid vortex has the properties of both global and local vortices \cite{Eto:2009kg}.
The vortex tension logarithmically diverges in $r$ (in infinite space), but the color-magnetic flux
is well squeezed inside the vortex core.
Indeed, as shown in \cite{Eto:2009kg}, the profile functions $\{G(r),h(r)\}$ 
in Eqs.~(\ref{eq:ans_s}) and (\ref{eq:ans_a})
get exponentially small
of order ${\cal O}(e^{- m r})$ 
at a large distance 
$m r \gg 1$ with $m$ being $\min\{m_G,m_\chi\}$. 
This implies that the wave functions of the massless NG bosons $\mathbb{C}P^2$
are well localized inside the vortex core. 
We thus expect these modes are {\it normalizable}.
If it is the case, the NG modes propagate along the world-sheet of the color-flux tube.
The purpose of this section is to prove the normalizability and derive the $d=1+1$ dimensional effective theory
of the NG modes.

Before going to derivation of the effective action,
let us identify the $\mathbb{C}P^2$ zero modes in the background solutions. 
To this end, we take a singular gauge, $U=\exp\left(i\sqrt{\frac{2}{3}}T_8\theta\right)$, 
which transform the Ansatz (\ref{eq:ans_s}) and (\ref{eq:ans_a}) to other form 
\beq
\Phi^{\star} &=& v\,
e^{\frac{i\theta}{3}}
\left(
\frac{F(r)}{\sqrt 3} T_0 - \sqrt{2 \over 3} G(r) T_8
\right),
\\
A_i^\star &=& - \frac{1}{g_s} \frac{\epsilon_{ij}x^{j}}{r^2}
h(r) \sqrt{\frac{2}{3}} T_8, 
\eeq 
as keeping the topology unchanged. 
Starting from this special solution, 
the generic solutions can be obtained by acting the
color-flavor locked symmetry as
\beq
\Phi(U) \to U \Phi^\star U^{-1},\quad
A_i(U) \to U A_i^\star U^{-1},\quad
U \in SU(3)_{\rm C+F}.
\eeq
This action changes only $T_8$ 
with keeping $T_0$. 
We define coordinates on $\mathbb{C}P^2$ by 
\beq
- U \left(\sqrt{\frac{2}{3}}T_8\right) U^{-1} \equiv   \phi  \phi^\dagger - \frac{{\bf 1}_3}{3}
\equiv \left< \phi \phi^\dagger\right>, \label{eq:phi}
\eeq
where $\phi$ is a complex $\NC=3$-column vector, 
and $\left< A \right>$ denotes the traceless part 
of a square matrix $A$. 
The $SU(3)_{\rm C+F}$ symmetry acts on $  \phi$ from the left hand side as $  \phi \to U   \phi$.
Taking trace of this, one gets
\beq
 \phi^\dagger   \phi = 1.
\eeq
In the definition of $\phi$ in Eq.~(\ref{eq:phi}), 
there is a redundancy in the overall phase of $\phi$.
This brings us a $U(1)$ equivalence relation, $ \phi \sim e^{i\sigma}  \phi$, 
and therefore one finds that 
$\phi$ are indeed the (homogeneous) coordinates on $\mathbb{C}P^2$.

In order to derive the low-energy effective theory, 
we now promote the moduli parameters 
to the fields depending on the coordinates $x^\alpha$
with $\alpha = 0,3$
of the vortex world-sheet 
using the moduli space approximation 
(first introduced by Manton for BPS monopoles \cite{Manton:1981mp}
), 
$ \varphi \to  \varphi(x^\alpha)$.
We are interested in the slow deformation such as $|\p_\alpha  \varphi(x^\alpha)| \ll \min\{m_{\phi,\chi,G}^{-1}\}$.
From a symmetry argument 
the low-energy effective theory on the world-sheet can be written 
in the form of the ${\mathbb C}P^2$ (non-linear sigma) model,  
\beq
{\cal L}_{\rm low} = C\,g_{ab^*}(\varphi,\varphi^*) 
  K_{\alpha} \p_\alpha \varphi^a \p^\alpha \varphi^{b*},\qquad
 (a,b=1,2)
\label{eq:eff_formal}
\eeq
with 
the so-called Fubini-Study metric 
$g_{ab^*} = (\delta_{ab}(1+|\varphi|^2)- \varphi^{*a}\varphi^b)/(1+|\varphi|^2)^2$ on the complex projective space ${\mathbb C}P^2$. 
The overall coefficient $C$ 
 in front of the Lagrangian is a certain real constant 
 (corresponding to the pion decay constant in the chiral Lagrangian).
It is called the K\"ahler class in the context of 
the K\"ahler geometry.
It should be calculated from the Ginzburg-Landau theory, and consequently depends on the parameters $\{m_\chi,m_\phi,m_G\}$. 
If $C$ is finite (infinite) the NG modes are (non-)normalizable. 
We find $C$ to take finite values in explicit calculations for various 
parameters $\{m_\chi,m_\phi,m_G\}$, below.

The effective Lagrangian (\ref{eq:eff_formal}) in the quadratic order of the derivatives $\p_\alpha$ ($\alpha=0,1$) can be obtained from the
original Lagrangian (\ref{eq:lsm}) in the following procedure. 
We substitute the background solution, where the orientation modes are 
promoted to the fields, into the original Lagrangian 
(\ref{eq:full_lag}), to yield
\beq
{\cal L}_{\rm low} = \int dx^1dx^2\ 
\Tr\left[- \frac{1}{2} F_{i\alpha}F^{i\alpha} 
+ K_\alpha  \D_\alpha \Phi^\dagger\D^\alpha \Phi 
+ K_0 (\alpha_1 j_0 + \alpha_2 \tilde j_0)\right],\qquad (\alpha=0,3)
\label{eq:eff}
\eeq
with $\Phi = \Phi(\phi(x^\alpha)),\ A_m=A_m(\phi(x^\alpha))$.
Note that the $x^\alpha$-dependence appears in the Lagrangian only through the moduli fields $ \phi(x^{\alpha})$.
We have already known the $x^\alpha$-dependence of $\Phi$ and $A_{i=1,2}$ 
\beq
\Phi(r,\theta, \phi(x^\alpha)) &=& v\, e^{\frac{i\theta}{3}} \left(
\frac{F(r)}{\sqrt 3}T_0 + G(r) \left< \phi (x^\alpha) \phi^\dagger (x^\alpha) \right>
\right),\\
A_i(r,\theta, \phi(x^\alpha)) &=& \frac{1}{g_s} \frac{\epsilon_{ij}x^{j}}{r^2}
h(r) \left< \phi (x^\alpha) \phi^\dagger (x^\alpha)\right>.
\eeq

The missing piece to construct the low-energy theory is $A_{\alpha}( \phi(x^\alpha))$ which is zero in the 
background configurations, namely it does not depend on neither $x^1$ nor $x^2$,
but does not vanish for fluctuations. 
We make an ansatz for $A_\alpha$ by following \cite{Gorsky:2004ad}
\beq
A_\alpha(\phi(x^\alpha)) = \frac{i\rho(r)}{g_s} \left[ \left< \phi \phi^\dagger\right> , 
\p_\alpha\left< \phi \phi^\dagger\right>\right],
\eeq
where 
$\rho(r)$ is an unknown function which will be determined below.
In order to make the following calculations simplified, let us define 
\beq
{\cal F}_\alpha(a,b) \equiv 
a \phi \p_\alpha{\phi}^\dagger + b \p_\alpha\phi\phi^\dagger 
+ (a-b) \phi\phi^\dagger\p_\alpha\phi\phi^\dagger,\quad a,b \in \mathbb{C}.
\eeq
One finds that this quantity satisfies the relations 
\beq
 && {\cal F}(a,b)^\dagger = {\cal F}(b^*,a^*),\quad  \Tr [{\cal F}(a,b)]=0,\\
 && \alpha {\cal F}(a,b) = {\cal F}(\alpha a,\alpha b), \quad
    {\cal F}(a,b) + {\cal F}(a',b') = {\cal F}(a+a',b+b') ,\\ 
 && \Tr\left[{\cal F}_\alpha(a,b)^\dagger {\cal F}^\alpha(a,b)\right] = 
(|a|^2 + |b|^2) 
[
 \p^\alpha\phi^\dagger \p_\alpha \phi 
 + (\phi^\dagger \p^\alpha\phi) (\phi^\dagger \p_\alpha\phi)] .
\eeq
The effective Lagrangian (\ref{eq:eff}) consists of the terms 
$\D_\alpha \Phi$ and $F_{i\alpha}$ 
which include $\p_\alpha$ once and are traceless, so that 
they can be written in terms of ${\cal F}(a,b)$:
\beq
A_\alpha &=& \frac{i\rho}{g_s} {\cal F}_\alpha(1,-1),\\
 \D_\alpha \Phi 
 &=& v\, e^{\frac{i\theta}{3}} {\cal F}_\alpha(f-g +\rho g,f-g -\rho f),\\
 F_{\alpha i} &=&  \frac{1}{g_s}\epsilon_{ij}
 \frac{x_j}{r^2}\,(1-\rho)h\, {\cal F}_\alpha(1,1) 
 - \frac{i}{g_s}\frac{x_i}{r^2}\rho' {\cal F}_\alpha(1,-1).
\eeq
By plugging these into Eq.~(\ref{eq:eff}), we finally obtain
\beq
{\cal L}_{\rm low} = C\, \sum_{\alpha=0,3} 
 K_\alpha [
 \p^\alpha\phi^\dagger \p_\alpha \phi 
 + (\phi^\dagger \p^\alpha\phi) (\phi^\dagger \p_\alpha\phi)],
\label{eq:low_lag}
\eeq
where we have defined $K_3 \equiv K_1$, 
and the constant $C$ is given by the integration 
\beq
C = \frac{4\pi}{g_s^2} \int dr\ \frac{r}{2}
\left[
m_G^2\left( 
(1-\rho)(f-g)^2 + \frac{\rho^2}{2}(f^2+g^2)
\right)
+ \frac{(1-\rho)^2 h^2 }{r^2} + \rho'{}^2
\right].
\label{eq:kahler_class}
\eeq
Note that $j_0$ and $\tilde j_0$ 
in the original Lagrangian (\ref{eq:full_lag}) give 
no contribution in the low-energy effective action because of 
the equation
\beq
\Tr[\Phi^\dagger \D_\alpha\Phi] = v^2 G\, \Tr\left[\left<\phi\phi^\dagger\right> 
{\cal F}_\alpha(f-g +\rho g,f-g -\rho f) \right] = 0.
\eeq

The unknown function $\rho(r)$ should be determined in such a way
that the ``Hamiltonian", Eq.~(\ref{eq:kahler_class}), 
is minimized.
The Euler-Lagrange equation of motion for $\rho$ read 
\beq
\rho'' + \frac{\rho'}{r} + (1-\rho) \frac{h^2}{r^2} - \frac{m_G^2}{2} \left[(f^2+g^2)\rho - (f-g)^2\right]=0.
\label{eq:rho}
\eeq
We have to solve this with given background solutions $\{f,g,h\}$ and the boundary condition
\beq
\rho \to
\left\{
\begin{array}{ccl}
1 & & \text{for}\quad r\to0,\\
0 & & \text{for}\quad r\to\infty.
\end{array}
\right.
\eeq
Note that the K\"ahler class $C$ expressed in terms of $\rho$ 
in Eq.~(\ref{eq:kahler_class}) and the Euler-Lagrange equation
(\ref{eq:rho}) are formally the same as those 
for the BPS non-Abelian local vortex 
in the supersymmetric $U(N)$ gauge theory \cite{Gorsky:2004ad}.
Eq.~(\ref{eq:rho}) for the BPS non-Abelian vortex \cite{Gorsky:2004ad} has been analytically solved to give $\rho =1-f/g$
with an aid of the supersymmetry. 
Since the equations (\ref{eq:1}) and (\ref{eq:2}) 
for $f,g$ in the present case 
are different from the BPS equations 
in the supersymmetric theory,  
Eq.~(\ref{eq:rho}) cannot be solved analytically. 

\begin{figure}
\begin{center}
\includegraphics[width=17cm]{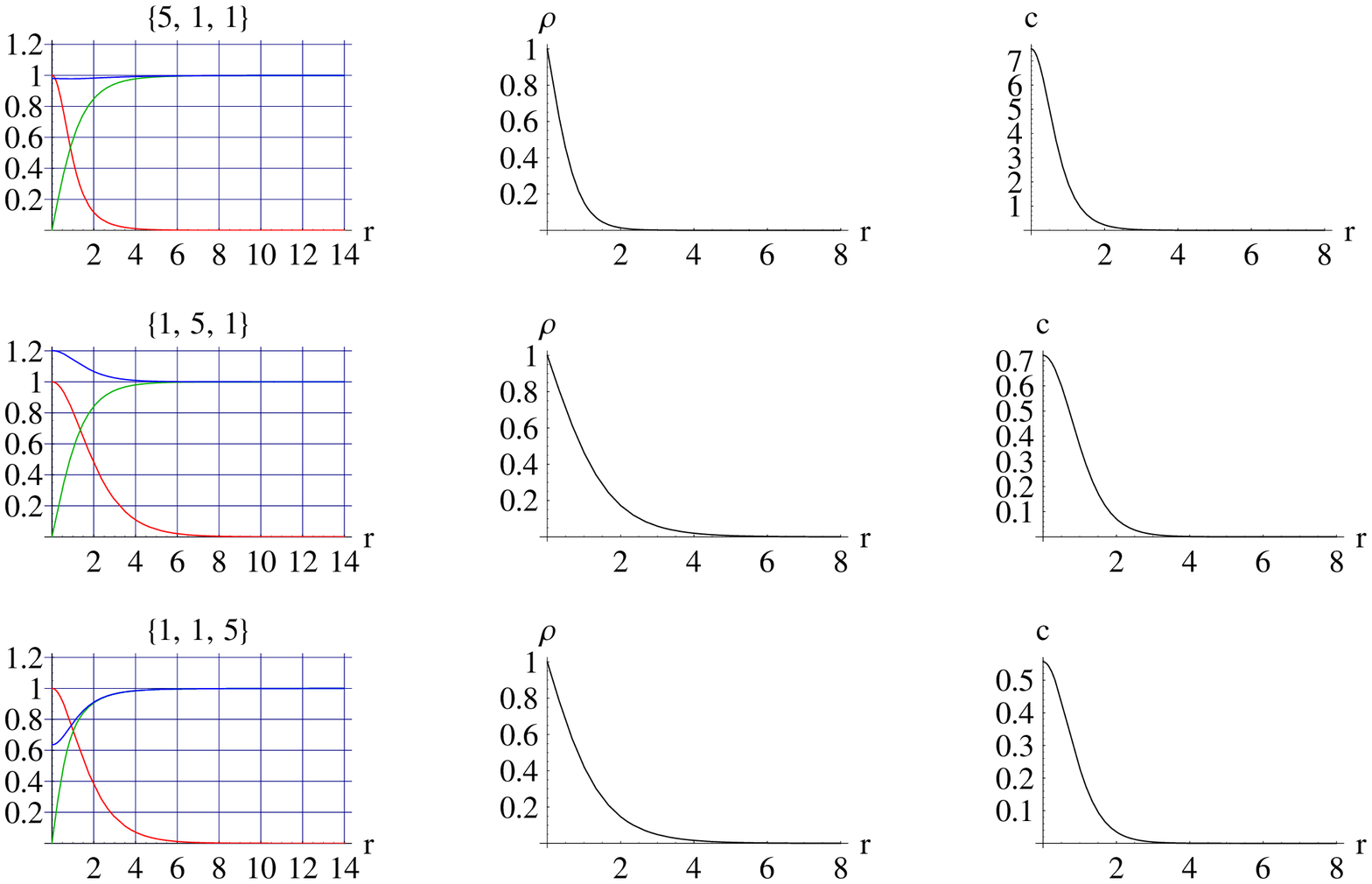}
\includegraphics[width=17cm]{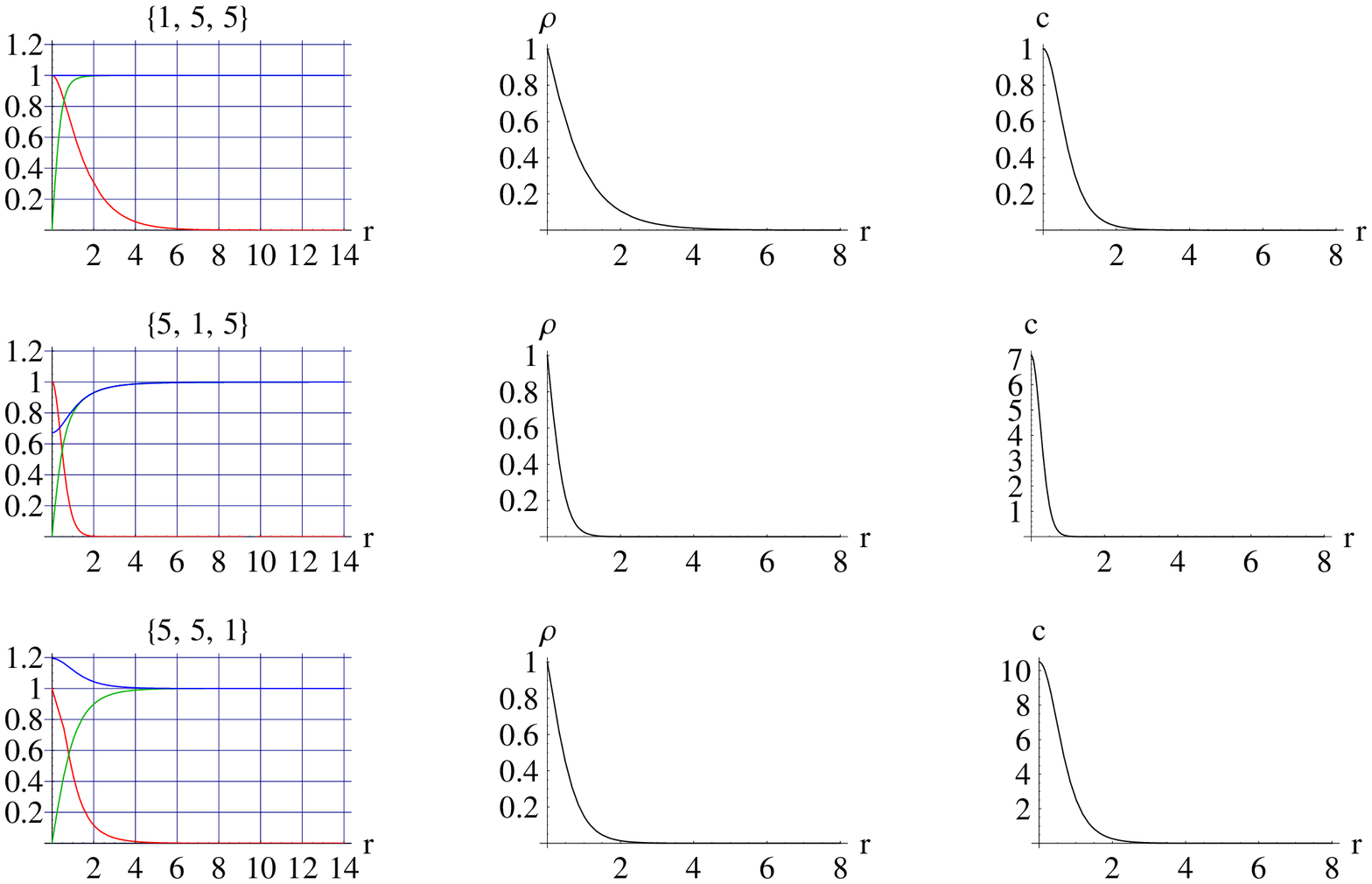}
\caption{{\footnotesize 
The background configurations with various $\{m_G,m_\phi,m_\chi\}$ 
are shown in the left panels. 
The middle panels show the function $\rho(r)$, 
and the integrand $c$ ($C = \frac{4\pi}{g_s^2}\int dr\ r c$) 
in Eq.~(\ref{eq:kahler_class}) is 
shown in the right panels.}}
\label{fig:1}
\end{center}
\end{figure}

In order to solve it, we first need specify the
background configurations $f,g,h$ by solving Eqs.~(\ref{eq:1})-(\ref{eq:3}) \cite{Eto:2009kg}. 
Then we numerically solve Eq.~(\ref{eq:rho}) with the background fields.
Various numerical solutions are shown in Fig.~\ref{fig:1}.
Our numerical results for $C$ are listed in Table \ref{tab:1}. 
\begin{table}[ht]
\begin{center}
\begin{tabular}{c|cccccc}
$\{m_G,m_\phi,m_\chi\}$ & $\{5,1,1\}$ & $\{1,5,1\}$ & $\{1,1,5\}$ & $\{1,5,5\}$ & $\{5,1,5\}$ & $\{5,5,1\}$\\
\hline
$g_s^2 C/4\pi$ & 3.27 & 0.60 & 0.37 & 0.38 & 0.51 & 4.29
\end{tabular}
\caption{{\footnotesize The K\"ahler class given 
in Eq.~(\ref{eq:kahler_class}).
The ratio $C/C_{\rm BPS}$ to the BPS case $C_{\rm BPS} = 4 \pi/g_s^2$ 
are written. 
}}
\label{tab:1}
\end{center}
\end{table}
We thus have shown that the K\"ahler class $C$ is finite for a wide class of 
$\{m_G,m_\phi,m_\chi\}$, 
implying that the massless  NG modes $\mathbb{C}P^2$ are normalizable 
on the world-sheet of the vortex.
Comparing to $C_{\rm BPS} = 4\pi/g_s^2$ for 
a local BPS non-Abelian vortex-string \cite{Gorsky:2004ad}, 
$C$ can be larger or smaller than it in general, 
depending on parameters.

\begin{figure}[ht]
\begin{center}
\includegraphics[height=5cm]{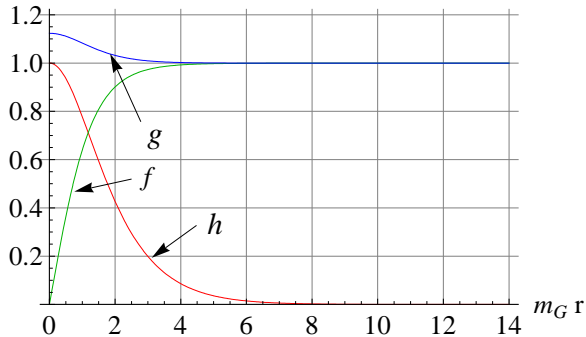}
\caption{{\footnotesize The vortex profile functions $\{f,g,h\}$ for 
$m_G \simeq 130$ MeV, $m_\phi \simeq 344$ MeV and $m_\chi \simeq 174$ MeV.
}}
\label{fig:qcd}
\end{center}
\end{figure}

Let us estimate the K\"ahler class $C$ in 
a realistic setting in the weak coupling regime, 
where the couplings of the Ginzburg-Landau Lagrangian have
been determined \cite{Iida:2000ha,Giannakis:2001wz}
as $\lambda_1 = \lambda_2 = 3 K_1 = \frac{7\zeta(3)}{4(\pi T_c)^2} N(\mu)$
and $m^2 = - 8 N(\mu) \log\frac{T}{T_c}$ with $N(\mu) = \frac{\mu^2}{2\pi^2}$.
We consider the quark chemical potential $\mu = 500$ MeV, $\Lambda = 200$ MeV,
$T_c = 100$ MeV and $T = 0.9 T_c$. Then we get
$g_s= \sqrt{\frac{12\pi^2}{(\frac{11}{2} \NC - \NF)\log\frac{\mu}{\Lambda}}} \simeq 3.1$,
$v\simeq 70$ MeV, 
$m_G \simeq 130$ MeV, $m_\phi \simeq 344$ MeV and $m_\chi \simeq 174$ MeV.
The numerical solution is shown in Fig.~\ref{fig:qcd} and we get $C = 0.503 \times \frac{4\pi}{g_s^2}$
given in Eq.~(\ref{eq:kahler_class}). 
We thus have found that the K\"ahler class $C$ in 
this realistic setting is about a half of $C_{\rm BPS}$ of 
the BPS case.

We see that the speed of NG modes propagating 
along a vortex-string is given by 
\beq
 v_c^2 = K_1/K_0 
\eeq
as expected from the original Lagrangian (\ref{eq:full_lag}).
Although we have started from the Lagrangian (\ref{eq:full_lag}) 
which has only the $SO(3)$ rotational symmetry of the space (without the Lorentz invariance), 
we have eventually arrived 
at the low-energy effective Lagrangian (\ref{eq:low_lag}) 
which has the effective Lorentz symmetry 
on the world-sheet
if we rescale $x^3 \to {x^3}' =v_c x^3$.

\section{Conclusion and Discussion} \label{sec:4}

We have derived the low-energy effective action 
for the orientational modes ${\mathbb C}P^2$ of 
a non-Abelian semi-superfluid vortex-sting 
in the CFL phase, 
and have confirmed that those modes are 
in fact normalizable and localized around the vortex-string.
The K\"ahler class has been evaluated on 
the background vortex solutions  
with various choices of the parameters (Table \ref{tab:1}).
It has been shown to be different from the one for 
a local BPS non-Abelian vortex-string but in general 
to be larger or smaller than it depending on parameters.

Our work will become the first step to study dynamics 
of semi-superfluid vortex-strings. 
When well-separated vortices constitute 
a lattice by a long range repulsion \cite{Nakano:2007dr}, 
the ${\mathbb C}P^2$ waves (as well as Kelvin waves) 
propagate along each vortex-string independently.  
Such waves will arise at finite temperature 
or when two vortex-strings reconnect 
as in Fig.\ref{fig:reconnection}.
It has been shown \cite{Eto:2006db} 
in the case of local non-Abelian vortices 
that the reconnection always occurs  
when two vortex-strings collide 
even if they have different orientations initially.

In study of those dynamics, 
we also have to include the interaction of a vortex-string 
with the $U(1)_{\rm B}$ Nambu-Goldstone mode(phonon) 
living in the bulk.
The string radiates or absorbs those particles because 
it is a source of them \cite{Vilenkin}.
This interaction can be written 
as in the same manner with the Abelian case, 
\beq 
S_{\rm int.} = 
 2 \pi \int d \sigma^{\mu\nu}B_{\mu\nu}
\eeq
where the $U(1)_{\rm B}$ NG boson has been dualized 
to the 2-form field $B_{\mu\nu}$. 
On the other hand it is an open question 
if the non-Abelian semi-superfluid vortex 
 interacts with the CFL mesons, 
the NG bosons for the broken chiral symmetry.

Our result can be used when vortices are well-separated 
compared with Compton wave lengths of massive particles. 
If two or more vortices are close to each other, 
we have to construct the effective action 
from the multiple vortex background. 
The construction of the effective action for 
multiple non-Abelian vortices was formally achieved 
in the BPS case in supersymmetric theories \cite{Eto:2006uw}.
When two vortices make a bound state 
the orientational zero modes 
are not the direct product of two ${\mathbb C}P^{N-1}$'s but 
something different \cite{Eto:2006cx}. 
Extensions of the present work to the multiple non-Abelian vortices 
at a short distance 
as well as a gas of non-Abelian vortices 
at finite temperature \cite{Eto:2007aw}
remain as interesting problems.

In the present work we have considered an ideal CFL phase 
where the exact flavor symmetry has held. 
Once flavor asymmetries in electric charges or mass differences are taken into consideration, 
there would appear favored directions 
in the ${\mathbb C}P^2$ space of the orientational zero modes, 
hence the effective action which 
we have derived here would be modified accordingly.  
We will discuss this problem elsewhere.

Finally we give a comment on possible application to instantons.
Instantons cannot stably exist but shrink to zero in the Higgs phase, 
due to the Derrick's scaling argument.
Instead, they can live stably inside a non-Abelian vortex core
where they are regarded as sigma model instantons 
in the ${\mathbb C}P^{N-1}$ world-sheet theory of the vortex.
In the case of supersymmetric QCD,  
the instanton energy(action) can be calculated 
as the lump energy(action) multiplied by 
the decay constant (K\"ahler class) of ${\mathbb C}P^{N-1}$ 
\cite{Hanany:2004ea},
because the energy of BPS solitons coincides with 
their topological charge in supersymmetric theories.
In our case of high density QCD, however, 
this agreement does not hold 
but the instanton energy inside a non-Abelian vortex 
becomes smaller or larger than 
the standard instanton energy. 
Physical interpretation of this phenomenon remains 
as a future problem.

\section*{Acknowledgement}

We would like to thank Naoki Yamamoto for a lot of useful comments.
The work of M.E. is supported by Special Postdoctoral Researchers Program at RIKEN.
The work of M.N.~is supported in part by Grant-in-Aid for Scientific
Research (No.~20740141) from the Ministry
of Education, Culture, Sports, Science and Technology-Japan.


\end{document}